# Reconfigurable integrated waveguide meshes for photonic signal processing and emerging applications

Daniel Pérez[a], Ivana Gasulla[a], José Capmany*[a]
[a]ITEAM Research Istitute, Universitat Politècnica de València, Camino de Vera s/n, 46022 València, Spain;

## ABSTRACT

We review the recent advances reported in the field of integrated photonic waveguide meshes, both from the theoretical as well as from the experimental point of view. We show how these devices can be programmed to implement both traditional signal processing structures, such as finite and infinite impulse response filters, delay lines, beamforming networks as well as more advanced linear matrix optics functionalities. Experimental results reported both in Silicon and Silicon Nitride material platforms will be presented. We will also discuss the main programming algorithms to implement these structures and discuss their applications either as standalone systems or as part of more elaborated subsystems in microwave photonics, quantum information and machine learning.

## 1. INTRODUCTION

Programmable Multifunctional Photonics (PMP) [1-8] is a new paradigm that aims at designing common integrated optical hardware configurations, which by suitable programming can implement a variety of functionalities that, in turn, can be exploited as basic operations in many application fields. PMP is therefore a transversal concept inspired in others, which are already employed in other technology fields. For instance in electronics, Field Programmable Gate Array (FPGA) devices enable a much more flexible universal operation as compared to Application Specific integrated Circuits (ASICs). In communications, Software Defined Networks (SDN) enable the exploitation and reconfiguration of a common set of resources provided by a network hardware infrastructure to ensure an optimum configuration in demand of time-varying requirements set upon several quality and bandwidth performance indicators. Finally, in wireless transmission Software Radio (SR) allows the emulation of different specific radiofrequency receivers with a single hardware platform. In the area of Photonics, the PMP approach aims to provide a complementary approach to that based on Application Specific Integrated Photonics Circuits (ASPICs). The objective is to leverage on the universal properties of this approach and seek similar advantages as FPGAs bring over ASICs in electronics as listed in table 1.

The current worldwide interest in PMP is justified by the surge of a considerable number of emerging applications that are and will be calling for true flexibility, reconfigurability and low-cost, compact and low-power consuming devices. In the field of telecommunications PMP can be instrumental in a series of functionalities, such as the implementation of arbitrary mode converters [9,10], fiber-wireless interfacing devices [11] and broadband switches [12], which can also form the basis for computer interconnection [13]. In the field of sensing, PMP can lead to a generic class of programmable measuring devices [14], which might be successfully integrated as a building block in the future Internet of Things (IoT). In quantum information technologies, PMP can open the path to large-scale quantum gates and circuits based on unitary matrix transformations [15-19], a feature that can also be exploited in reconfigurable neurophotonic systems and Fourier based optical signal processors [20].

All in all, the success of PMP relies on the research of a suitable interconnection hardware architecture that can offer a very high spatial regularity as well as the possibility of independently setting (with a very low power consumption) the interconnection state of each connecting element. Integrated waveguide meshes provide regular and periodic geometries, formed by replicating a unit cell, which can take the form of a square, hexagon or triangle, among other configurations. Each side of the cell is formed by two integrated waveguides connected by means of a Mach-Zehnder Interferometer (MZI) that can be operated by means of an output control signal as a crossbar switch or as a variable coupler with independent power division ratio and phase shift. A mesh formed by a suitable amount of unit cells can be programmed to implement a wide variety of functionalities much in the same mood as a FPGAoperates in electronics [21].

Table 1. Basic features of ASIC and FPGA approaches in electronics [21].

|  | ASIC | FPGA |
|---|---|---|
| **Time to Market** | Slow | Fast |
| **Non-Recurring Engineering** | Very High | Low |
| **Unit Cost** | Low | Medium |
| **Design Flow** | Complex | Simple |
| **Performance** | High | Medium |
| **Application Flexibility/Versatility** | Very low | High |
| **Power consumption** | Low | High |
| **Size** | Low | Medium |

In this paper, we review the recent advances reported in the field of integrated photonic waveguide meshes, both from the theoretical as well as from the experimental point of view. In section 2, we provide a review of the integrated waveguide mesh concept including the description of its basic configuration element, the Tunable Basic Unit (TBU), which is implemented by means of a MZI and we show how TBUs can be programmed to operate in a cross/bar or as a tunable coupling device providing independent amplitude and phase values. In section 3, we review how waveguide meshes can be programed to implement both traditional signal processing structures, such as finite and infinite impulse response filters, delay lines, beamforming networks, as well as more advanced linear matrix optics functionalities. Section 4 discusses some experimental results reported both in Silicon and Silicon Nitride material platforms and outlines some of the practical challenges to be overcome in future designs. In section 5, we provide the main programming algorithms to implement these structures, in particular some detailed discussion on the algorithms that can be programmed to implement arbitrary matrix transformations between the input and output waveguide ports, and discuss their applications either as standalone systems or as part of more elaborated subsystems in microwave photonics, quantum information, optical signal processing and machine learning. Finally, section 6 provides a summary wrap-up, some conclusions and directions for future work.

## 2. INTEGRATED PHOTONIC WAVEGUIDE MESHES

### 2.1 Basic architecture concept

A 2D integrated waveguide mesh is a structure where a unitary waveguide cell is spatially replicated [4],[5]. Each side of the unitary cell is composed of two integrated waveguides that are coupled by means of a TBU, the core of which is a balanced MZI. The application of a control signal to the MZI allows the independent amplitude and phase control of the signals coupled between the two waveguides as explained in the following subsection. Figure 1 shows the three main designs for waveguide meshes that allow both forward and backward propagation proposed to date: the *square*, the *triangular* and the *hexagonal* configurations. By suitably acting over the MZIs of each TBU, different propagation and interconnection paths can be established between one or several input/output ports.

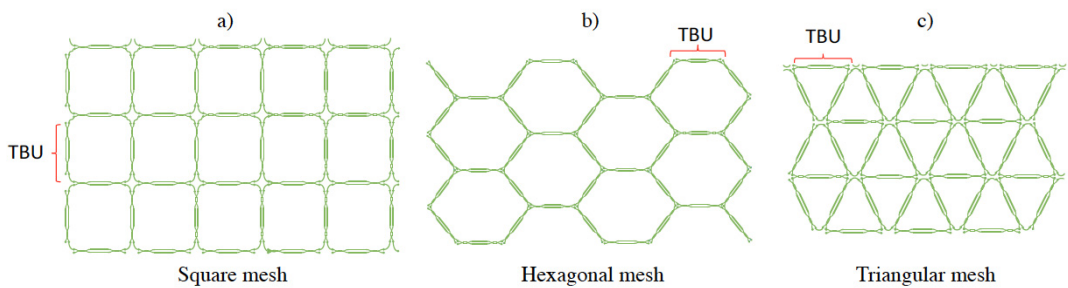

Figure 1. Several designs for 2D integrated waveguide meshes: (a) Square, (b) Hexagonal, (c) Triangular

In these topologies, 2D meshes allow both feedforward and feedbackward propagation of light. By means of external electronic control signals, each MZI in the mesh can be configured to operate as a directional coupler or simply as an optical switch in a cross or bar state providing independent amplitude- and phase-controlled optical routing. In this way, the combination of different MZIs in the 2D grid, -each individually configured as desired-, enables, in principle, the synthesis of any kind of optical core circuit topology, including finite and infinite impulse response filters, where the sampling period can be discretely tuned by appropriate switching along the beamsplitter-based squared mesh. According to recent studies, the hexagonal mesh is potentially the most flexible candidate alternative for implementing the PMP concept.

**2.2 The Tunable Basic Unit (TBU)**

The central basic element in the waveguide mesh is the TBU [5], [7]. Figure 2 (a) shows in its upper part the block layout of the TBU including its access and tunable coupling sections. The lower part shows a particular implementation of the TBU by means of a 3-dB balanced two heater-electrode MZI device.

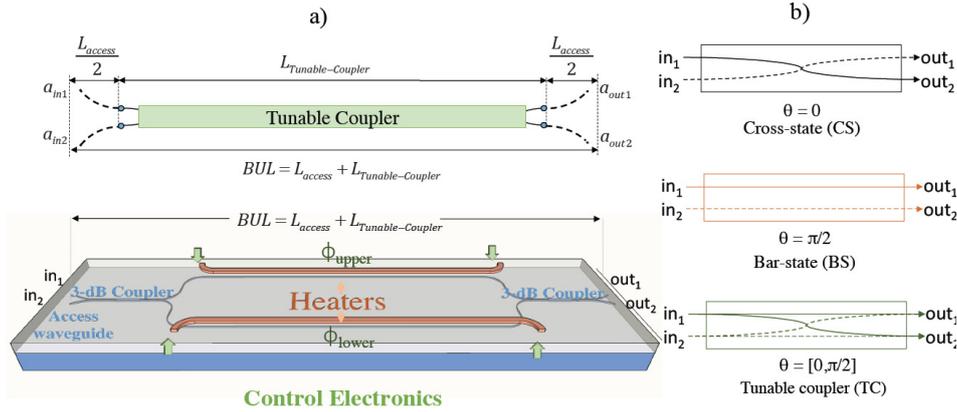

Figure 2. (a) Layout of a TBU, (b) TBU operation modes

The TBU is composed of the tunable coupler and its access (input/output) waveguides. The geometry of the latter is a function of the bending radius and varies for each mesh topology due to different angle between elements. The basic unit length (BUL) is BUL= $L_{access}$ + $L_{Tunable-Coupler}$, where $L_{access}$ is the overall length of the access waveguide segment and $L_{Tunable-Coupler}$ is the length of the tunable coupler.

**2.3 TBU Programming**

Referring to Fig. 2(b) the TBU can be programmed to implement 3 different states: cross state switch (light path connects $in_1$ to $out_2$ and $in_2$ to $out_1$), bar state switch (light path connects $in_1$ to $out_1$ and $in_2$ to $out_2$) and tunable splitter. For a balanced MZI loaded with heaters on both arms, the splitting ratio is obtained by increasing the effective index due to the Joule effect in the upper or lower arm, producing a $\phi_{upper}$ and $\phi_{lower}$ phase shift respectively. Once set, a common drive in both heaters will provide a common phase shift, leading to independent control of the amplitude ratio and the phase. The device matrix is defined by:

$$h_{TBU} = je^{j\Delta} \begin{pmatrix} \sin\theta & \cos\theta \\ \cos\theta & -\sin\theta \end{pmatrix} \gamma, \tag{1}$$

where $\theta$ is ($\phi_{upper}$- $\phi_{lower}$)/2 and $\Delta$ is ($\phi_{upper}$+ $\phi_{lower}$)/2. The coupling factor $K$ is then defined as $cos^2(\theta)$. Finally, $\gamma$ is a general loss term that includes the effect of propagation losses in the access waveguides, the tunable coupler waveguide and the insertion losses for both 3-dB couplers.

## 3. PROGRAMMING INTEGRATED WAVEGUIDE MESHES

**3.1 Waveguide mesh allocation in a general Photonic Processor**

While the 2D waveguide mesh is the key central element required in PMP there are other components that are needed to configure a programmable processor [11],[21]. The left part of Fig. 3 represents a general photonic processor architecture useful in a wide range of applications. All the elements are connected to the 2D reconfigurable photonic integrated waveguide mesh in such a way that, not only they produce the desired processing engines, but also connect the internal and the external elements required for different functionalities. The processor includes, as shown, both passive and active photonic components, interface ports with electronic control signals and RF driving input and output ports. Also pure input/output optical ports can be directly accessed

As highlighted in the right part of Fig. 3, a hybrid design might be needed to achieve the most efficient performance. In this case, one could choose a low loss silicon passive platform (ochre colour) for the passive devices and Indium Phosphide (red colour) for the active devices. Note that an array of optical amplifiers in this platform would be required to overcome the large conversion losses when moving from the radiofrequency to the optical domain. These losses are mainly related to the conversion efficiency of modulators and photodetectors as well as the propagation losses.

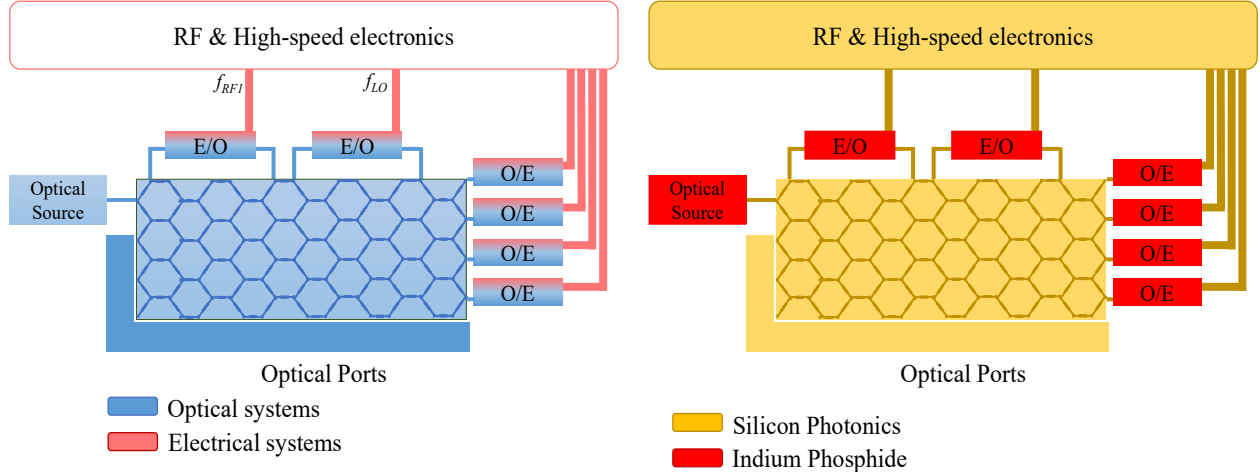

Figure 3. (Left) General photonic integrated processor architecture and (right) candidate fabrication platforms for each subsystem (after [21])

## 3.2 SISO and 2x2 input/output signal processing operation

The 2D integrated waveguide mesh can be programmed to operate as a standard Single Input/ Single output (SISO) or a 2x2 signal processor [4], [5], [7]. In principle, it has been demonstrated that internal connections set by proper biasing of the intermediate TBUs can enable the programming of a wide variety of functionalities including: Finite Impulse Response (FIR) filtering implemented either by 2x2 unbalanced MZI lattice structures and transversal filters, Infinite Impulse Response (IIR) Filters including single and multi-cavity resonators and more complex hybrid structures such as Coupled Resonator Optical Waveguides (CROWs) and Side-coupled Integrated Spaced Sequence of resonators (SCISSORs). Adequate programming also allows the implementation of tunable true time delay lines. As an example, Figure 4 shows the settings and programming of a hexagonal 2D integrated waveguide mesh to provides different CROW and SCISSOR structures [7].

## 3.3 MIMO and MxN input/output signal processing operation

A second and probably more versatile mode of operation is as a Multiple Input/Multiple output (MIMO) processor that implements an arbitrary unitary transformation. In essence, this task is equivalent to that of a linear optics device, which transforms a series of $N$ orthogonal modes ($|\phi_I\rangle$) into the corresponding N orthogonal modes at the output ($|\phi_O\rangle$) [1],[2], [8]. This transformation is defined by means of a unitary matrix $U$ ($|\phi_O\rangle = U|\phi_I\rangle$). Linear transformations are the fundamental building block of many applications in quantum information and communication systems, switching and routing, microwave photonics and optical channel management and supervision. The 2D hexagonal integrated waveguide mesh enables the implementation of the two layout versions of the universal linear interferometer proposed in the literature. The first case corresponds to a Reck-Miller triangular arrangement interferometer [22]. Figure 5 (a) displays an example of a 4 x 4 interferometer implemented by means of a triangular arrangement of beamsplitters and Fig. 5 (b) shows the equivalent structure implemented on a hexagonal waveguide mesh. Each beamsplitter can set a certain splitting ratio and a relative phase to the upper output. Reck et al. and Miller have developed algorithms to program and configure the triangular arrangement so it can implement any desired linear unitary transformation [22]. To adapt, for example, the synthesis algorithm developed by Miller to the hexagonal waveguide mesh we, first of all, need to consider the possible different phase contributions due to the different access paths established between the interferometer inputs and the internal processing elements forming the triangular arrangement of beam splitters and, from these, to the different outputs.

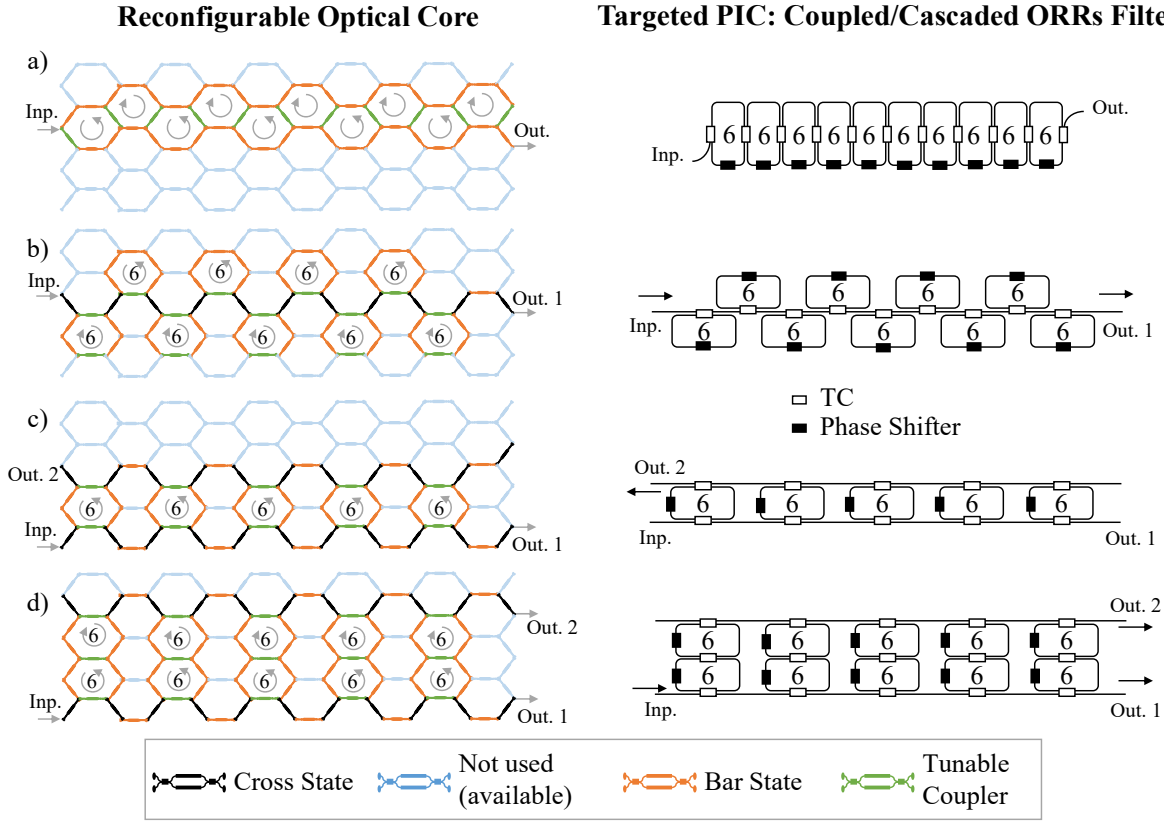

Figure 4. (Left) Settings for CROW & SCISSOR filter implementations in the hexagonal mesh core for (Right) (a) 10th-order CROW, (b) 9th-order Single channel SCISSOR, (c) 5th-order Double channel SCISSOR, and (e) twisted double channel SCISSOR.

These different phase contributions must be compensated. Then, we need to establish an equivalent configuration, -using the available elements in our hexagonal waveguide mesh-, to the MZI with a phase shifter in the upper output port employed by Miller and shown in Fig. 5(c). In our case, as illustrated in Fig. 5(d), the equivalent "beamsplitter" is implemented using a TBU for the tunable coupler (with a transfer matrix defined by $h_{TC}$ as in Eq. (1)), followed by two TBUs, which are biased in cross state and employed as output connections. In the latter, the upper TBU also implements a phase shifter and is defined by the transfer matrix $h_{UPS}$. The lower TBU is defined by the transfer matrix $h_{LPS}$. Miller's synthesis algorithm is based on writing any of the input basis functions as a linear combination of each input port or rectangular functions ($|\phi_{1n}\rangle$), and configuring sequentially each row of beam couplers for each input mode. These input modes can be obtained from the columns of the Hermitian Adjoint of the matrix $U$. A procedure describing the synthesis algorithm adaptation is discussed in section 5. The novel multiport interferometer configuration based on a rectangular arrangement proposed by Clements et al [23] can also be emulated using the hexagonal 2D integrated waveguide mesh. Figure 6 (a) displays an example illustrating the implementation of a 4 x 4 multiport interferometer. Figure 6(b) shows the equivalent structure implemented on a hexagonal waveguide mesh. In the algorithm developed by Clements et al. [23], each beamsplitter (Fig. 6(c)) sets a certain splitting ratio and a relative phase sequentially to program and configure the whole rectangular arrangement so it can implement any desired linear unitary transformation efficiently.

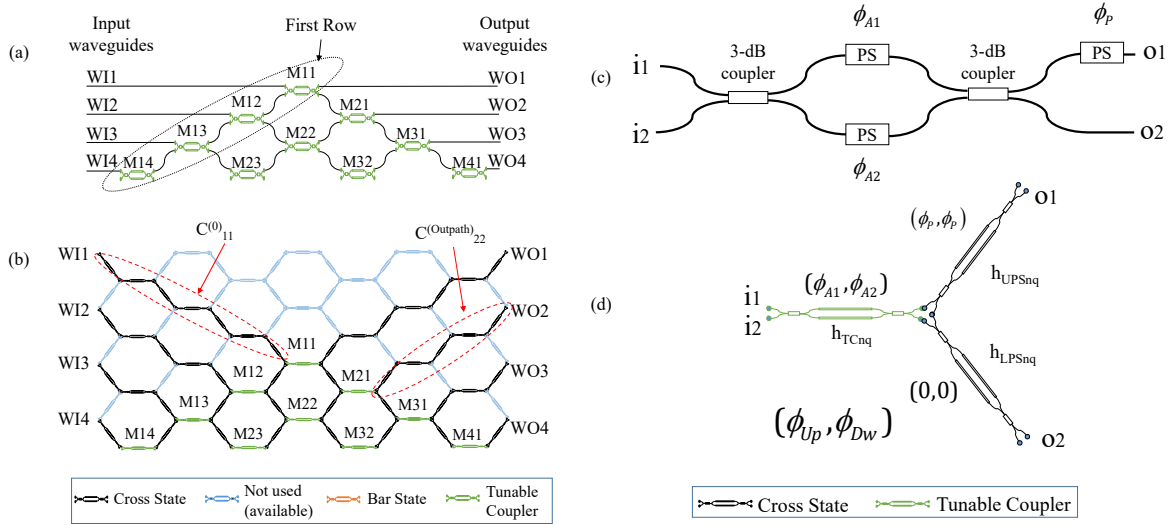

Figure 5. Universal interferometers: (a) Classical triangular arrangement and (b) hexagonal mesh based implementation of a 4 x4 interferometer. (c) Beamsplitter for the classical approach and (d) corresponding beamsplitter implementation with 3 TBUs for the hexagonal waveguide Mesh.

To adapt this layout and its synthesis algorithm to the hexagonal waveguide mesh, we need to perform a few modifications [8]. First of all, we must use a different matrix for the beam coupler/TBU structure. In our case, as can be seen in Fig. 6(d), we employ a TBU for the tunable coupler (colored in green), defined by a transfer function $h_{TC}$, and the two precedent TBUs (colored in black) for the required connections. Here, the upper one operates in cross mode providing an extra phase shifting (Upper Phase Shifter, $h_{UPS}$), while the lower one operates in cross mode. Note that both the classical and hexagonal approaches of rectangular arrangements need an extra phase shifter at each channel output that is required for certain applications, indicated as (*) in Fig. 6(b). Finally, some of the outer TBUs that build up the interferometer must be configured to be phase-transparent. A procedure describing the synthesis algorithm adaptation is discussed in section 5.

## 4. EXPERIMENTAL RESULTS

### 4.1 Silicon Photonics

Silicon photonics is one of the most attractive integration platform for programmable waveguide meshes since it allows high-volume fabrication as well as high integration densities due to its low refractive index contrast and moderate propagation losses among 0.5-2.5 dB/cm.

We recently reported the results of a waveguide mesh composed of 7 hexagonal cells (30 thermally-tuned TBUs) fabricated in Silicon on Insulator. The chip photograph is shown in Fig. 7(a). The device was fabricated at the Southampton Nanofabrication Centre at the University of Southampton. SOI wafers with a 220-nm thick silicon overlayer and a 3-µm thick buried oxide layer were used (for more details on fabrication and testing see [7]).

Despite the simplicity of the layout depicted in Fig.7(a), the 7-cell structure is capable of implementing over 100 different circuits for MWP filtering applications (basic MZI, FIR transversal filters, basic tunable ring cavities and IIR filters, as well as compound structures such as CROWs and SCISSORs), true time delay lines and optical coherent interferometry. The basic delay was 13.5 ps, given by a BUL of 975 µm and a group index of 4.18.

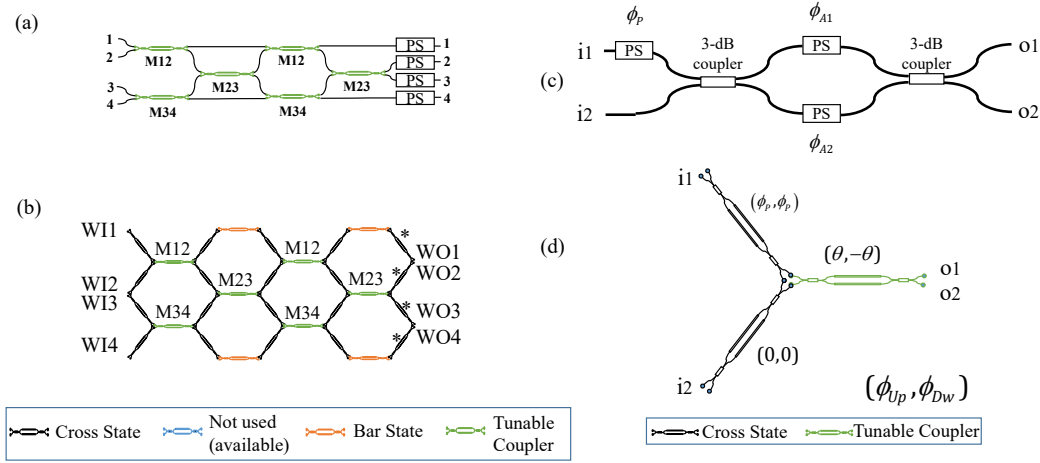

Figure 6. Universal interferometer: TBU settings of the hexagonal mesh for the rectangular arrangement implementation programming a Hadamard 4x4 linear transformation. ID: TBU identification Label, *K*: coupling constant, *ϕ*: additional phase shift, *P*: coupling constant when both phase shifters are unbiased. Green-color edges for TBUs acting as a tunable coupler, black color for the cross state and orange color for the bar state.

## 4.2 Silicon Nitride

Silicon Nitride waveguide meshes benefit from the integration of low propagation losses waveguides between 0.00045–1.5 dB/cm with moderate integration density values. Figure 7(c) shows the basic layout and photograph of the programmable optical chip architecture connecting thermally-tuned MZI devices in a square-shaped mesh network grid proposed by Zhuang and co-workers [4]. The structure, fabricated in $Si_3N_4$, comprised two square cells and was employed to demonstrate simple FIR and IIR impulse response filters with single and/or double input/output ports of synthetized ORRs. The processor featured a free spectral range (FSR) of 14 GHz. By appropriate programming of this processor, Zhuang et al. have demonstrated bandpass filters with a tunable center frequency that spans two octaves (1.6–6 GHz) and a reconfigurable band shape (including flat-top resonance with up to passband–stopband 25 dB extinction). They also demonstrated notch filters with up to 55 dB rejection ratio, Hilbert transformers and optical cavity-based tunable delay lines. The basic delay was greater than 19.7 ps, given by a BUL of 3450 μm and a group index of 1.72. We recently designed a mesh based on thermally-tuned 40 TBUs which is shown in Fig. 7(b). In this case, we re-designed the shape of the TBU to achieve a more compact layout and increase the component integration density. This chip has been fabricated in a Multi-Project Wafer run at a fabrication platform developed by VLC Photonics and the Centro Nacional de Microelectrónica (CNM) [24], and is currently under test. Based on previous and current fabrication runs we expect a maximum FSR of 60 GHz, given by a basic time delay of 8.42 ps (group index of 1.92 and a BUL of 1315 μm).

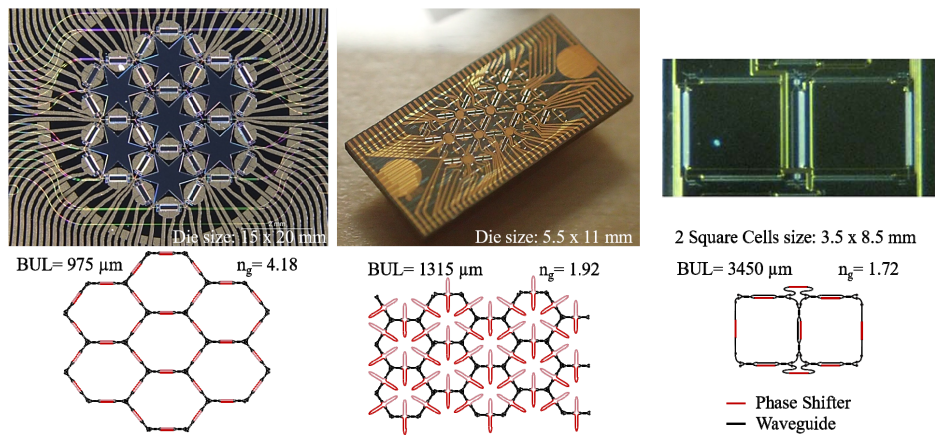

Figure 7. Chip picture and fabricated layout for different waveguide meshes: (a) Hexagonal topology in Silicon [7], (b) Hexagonal topology in SiN with modified TBU scheme [24], (c) Square topology in SiN [4].

## 4.3 Experimental Results

By suitably tuning the TBUs in the 7-cell hexagonal waveguide mesh, we have been able to program a wide variety of PIC topologies and design parameters [7],[8]. For example, Figure 8 illustrates a single cavity optical ring resonator with a cavity length given by 6 BULs. The figure shows in (a) the waveguide mesh configurations (with the TBU device status according to the color code previously described), (b) the circuit layout and (c) the modulus response for the OUT1 port. The measured results correspond to different values of $K_1$ and $K_2$, which settle the positions of the zero and the pole. The IIR filter tunability, which is shown in Fig. 8 (d), is achieved by exploiting the fact that the coupling constant and the phase shift in any TBU of the mesh can be adjusted independently. Hence, any TBU inside the cavity can be operated as a constant-amplitude phase shifter. Finally, Fig. 8 (e) shows the time-response of the ring resonator when the critical coupling is achieved.

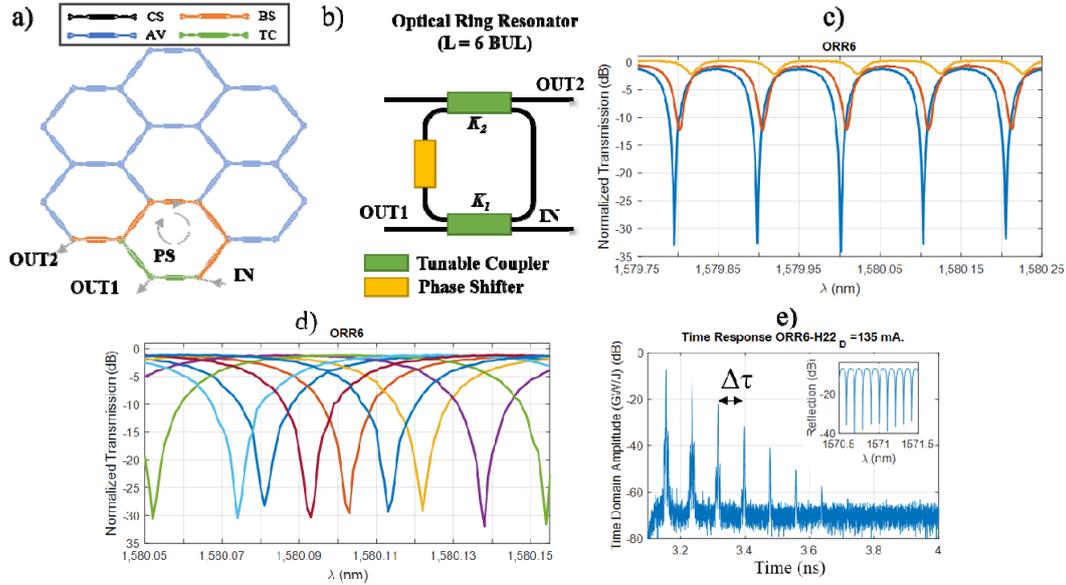

Figure 8. Experimental results for 6 BUL ring resonator IIR and FIR+IIR filters. (a) Waveguide mesh connection diagram, (b) circuit layout and (c) measured modulus transfer function for a IIR filter for different values of the coupling constants $K_1$ and $K_2$, (d) IIR filter along a full spectral period for different values of the optical ring resonator round-trip phase shift, (e) time response for critical coupling condition.

We programmed the 7-cell waveguide mesh to demonstrate several 3×3 and 4×4 linear unitary transformations. These are relevant examples of signal processing tasks that are needed in different applications. Figure 9 illustrates an experimental example of a 3x3 linear unitary transformations corresponding to a three-way beamsplitter and a Discrete Fourier Transform (DFT) configuration. In all cases, the measured results show an excellent agreement with the targeted matrices for the operation wavelength of $\lambda = 1580$ nm with an extinction ratio >25 dB between the 1 and 0 coefficients. The required values for the coupling constants and phases of the TBUs used in the former implementations were obtained by the synthesis algorithm adaptation. The resulting coefficients are translated into the required injected currents to the phase shifters according to the calibration curves obtained for each TBU during the chip characterization.

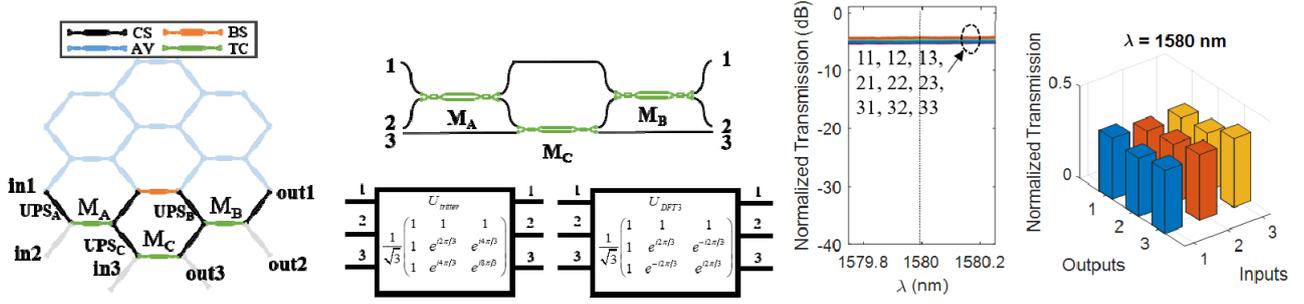

Figure 9. Unitary transformation configuration of a 3 x 43 interferometer based on a rectangular arrangement. (a) 7-cell configuration (CS = TBU in cross state, BS = MZI in bar state, TC = TBU in tunable coupler state, AV = TBU not employed), (b) circuit layout of the implemented interferometer, (c) spectral measurement of all input/output port connections, (d) normalized bar diagram of the resulting unitary matrix for λ = 1580 nm (after [8]).

### 4.4 Practical limitations

Ideally, a higher number of TBUs results in a more versatile waveguide mesh circuit. However, in practice, there exist footprint limitations together with several sources of degradation that must be considered: accumulated losses, imperfect coupling splitting ratios, phase control, parasitic back-reflections, loss imbalances, fabrication errors (gradients through the circuit in thickness or temperature), and drift in time, [11],[21].

Several works reporting the integration of a high-density MZI arrangement for matrix switching operations have succeeded at integrating more than 450 structures in a single die in a Silicon on Insulator platform, exceeding the Moore Law limits [26].

When designing programmable waveguide meshes, the designer faces an important miniaturization tradeoff: *minimum delay* and *accumulated losses*. The BUL and the group index will determine the minimum delay. For low refractive index difference platforms, the BUL is mainly limited by the tuning mechanism length and the 3-dB coupler lengths. 3-dB couplers in silicon can be reduced to less than 50 μm [27],[28], including the bend sections while the heaters can be reduced to 62 μm [29]. With the inclusion of bends and straight waveguides sections to increase the distance between both arms of the TBU to decrease thermal crosstalk, a total BUL of 240 μm seems potentially achievable. Assuming a typical SOI group index of 4.18, this is translated to maximum FSRs of around 150 and 50 GHz for the synthesis of MZIs and ORRs, respectively, in the hexagonal waveguide mesh topology. However, a reduction of the BUL implies that the signal must go through a greater number of TBUs to obtain a desired delay. If the 3-dB couplers limit the overall IL of the TBU, this miniaturization trade-off must be highly-considered. In fact, the main limitation of these structures resides in the number of accumulated losses. Assuming 0.1-dB loss 3-dB couplers, a path equivalent to 50 BULs will experience additional 10-dB losses than a classic waveguide of the same length.

Finally, the tuning mechanism impact on the final TBU length, tuning crosstalk and power consumption will limit the maximum number of active TBUs at the operational mesh and its performance. The use of alternative tuning mechanisms MEMS, piezoelectrics, or electromechanics are promising solutions to reduce the power consumption while enabling a reduction of the distance between the two TBU arms.

## 5. SYNTHESIS AND PROGRAMMING ALGORITHMS

### 5.1 Algorithms for SISO and 2x2 input/output

The available synthesis methods for the specific hardware SISO and 2x2 configurations that can be emulated using the 2D integrated waveguide mesh can be applied by developing a suitable procedure, which translates the results provided by the synthesis equations into specific parameter values of the TBUs that are needed to implement the waveguide coupling points in the emulated layout. This is possible for all the main FIR, IIR and combined FIR+IIR combination discrete-time signal processing hardware configurations employed in practice. For example, FIR filters are based either on cascades/lattices of 3-dB tunable MZIs or in transversal filter configurations. For both FIR filter alternatives, synthesis and recursive scaling algorithms have been developed in the literature [25] that are directly applicable since the hexagonal waveguide mesh can directly implement both 3 dB-tuneable MZI cascade lattices and transversal filter configurations. For IIR filters, either simple/compound optical ring cavities of ring-loaded 3-dB tuneable MZI cascades are employed. Again, synthesis algorithms have been reported in the literature [25] that are directly applicable since the hexagonal waveguide mesh can directly implement either simple or multiple cavity ring filters or ring-loaded 3-dB tunable MZI cascades.

## 5.2 Algorithms for MIMO and MxN input/output operation

The 2D integrated waveguide mesh can emulate as stated in section 3 the two configurations for universal interferometers reported in the literature. This means that the detailed synthesis and recursive scaling algorithms, which have been developed by Miller for triangular configurations [22] and by Clements et al. for rectangular [23] configurations can be applied provided that they are be adapted to the hexagonal waveguide mesh configuration. These adaptations have been reported in the literature [8],[21].

In a more general scope, the 2D integrated waveguide mesh can implement any Singular Value Decomposition (SVD) of an $M$x$N$ unitary matrix $D$. The SVD states that an $M$x$N$ matrix $D$ can be decomposed as [1]:

$$D = VD_{diag}U^{\dagger} \qquad (2)$$

where $U$ and $V$ are $N$x$N$ and $M$x$M$ unitary matrices and $D_{diag}$ is a $M$x$N$ diagonal matrix. Figures 10 and 11 illustrate a 4x4 and 5x5 SVD transformation using the 2D hexagonal integrated waveguide mesh and the triangular and rectangular interferometer configuration respectively.

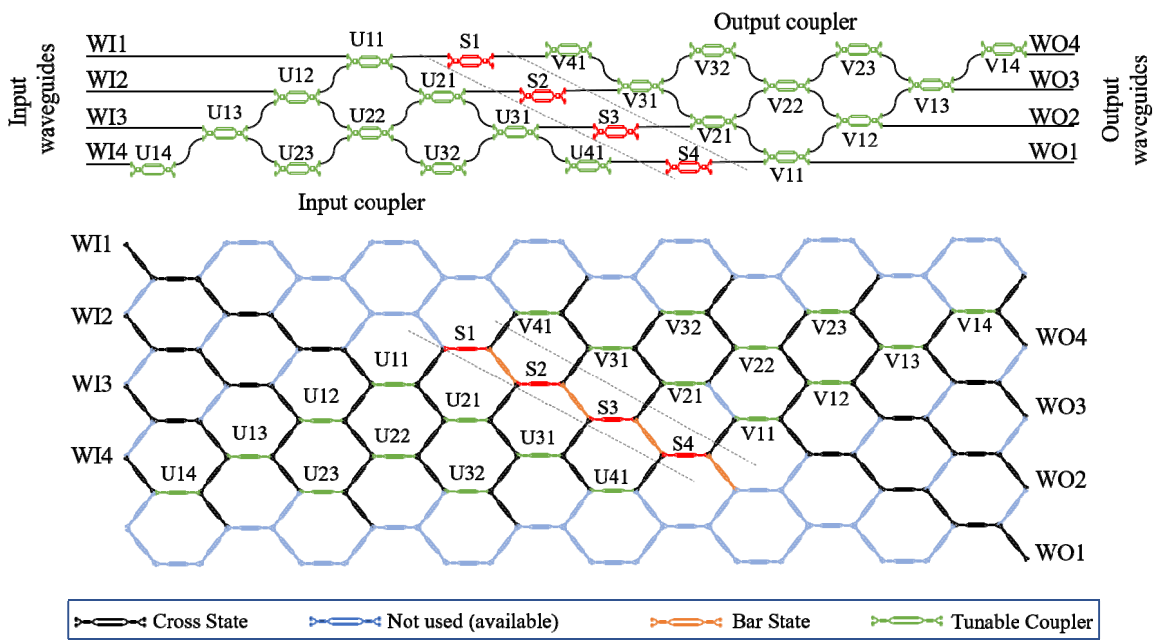

Figure 10. Triangular interferometer layout (Upper) and 2D hexagonal waveguide mesh implementation of a SVD for a 4x4 matrix transformation

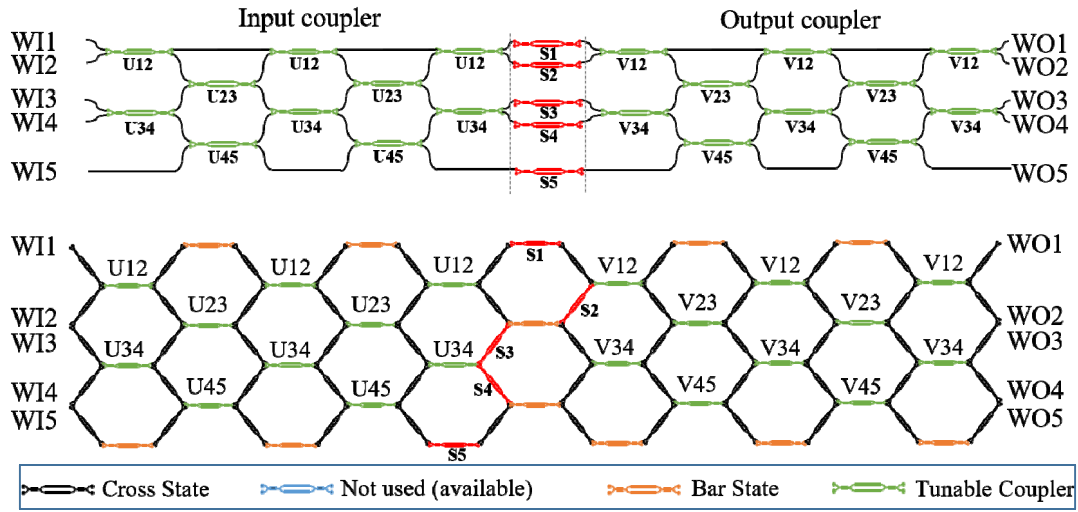

Figure 11. Rectangular interferometer layout (Upper) and 2D hexagonal waveguide mesh implementation of a SVD for a 5x5 matrix transformation

### 5.3 A review of Applications

PMP based on 2D integrated hexagonal waveguide meshes can find applications in a myriad of new emerging field as outlined in the introduction. In the field of telecommunications, we have, for instance provided a detailed discussion in [11] regarding their use in processor cores implementing the main required functionalities in Microwave Photonic systems and radio over fiber transmission. This configuration can also emulate a triangular or rectangular multiport interferometer, which could be employed for mode unscrambling in a similar way to the configuration reported in [10]. Another interesting field of application is switching and interconnections..

Hexagonal waveguide meshes emulating rectangular multiport interferometers can be programmed to manage different optical channels enabling broadcasting, add/drop configurations, multiplexing and demultiplexing functions to cite but a few. A different device can be obtained if we add optical ports, in not only the left and right side of the arrangement, but also in the top and bottom sides of the rectangular arrangement, a feature that is enabled by the hexagonal mesh topology. The main difference is that functionalities that are more compact can be achieved with this configuration. Consider that, for the standard rectangular arrangement, Add/drop functionality would require $N$ input ports equal to the number of channel inputs (I) and add channels (A). In the same way, the number of outputs ports will be equal to the number of output channels (O) and drop channels (D).

Figure 12 (top) illustrates this new configuration that places add and drop channels in the upper and bottom part respectively enabling a more efficient device. The TBUs from H1 to H5 are set in bar state performing the add/drop operation for the matrix illustrated in Figure 12 (top/right). In particular, corresponds to an Add (drop) operation for A1(D1) to A4(D4) while channel 5 bypasses the device. Figure 12 (lower) illustrates the fully reconfigurable interconnection matrix that can be programmed. Each block is a tunable coupler and can be configured as a switch or define a desired splitting ratio in order to enable broadcasting, multiplexing, demultiplexing, and switching operations.

The implementation of arbitrary unitary matrices enables the emulation of other applications, including quantum linear transformations, Fast Fourier Transforms and linear transformations preceding nonlinear threshold operation in neural networks and machine learning systems. Recent works reported on these fields are based on static hardware configurations, which can be emulated by 2D integrated waveguide meshes.

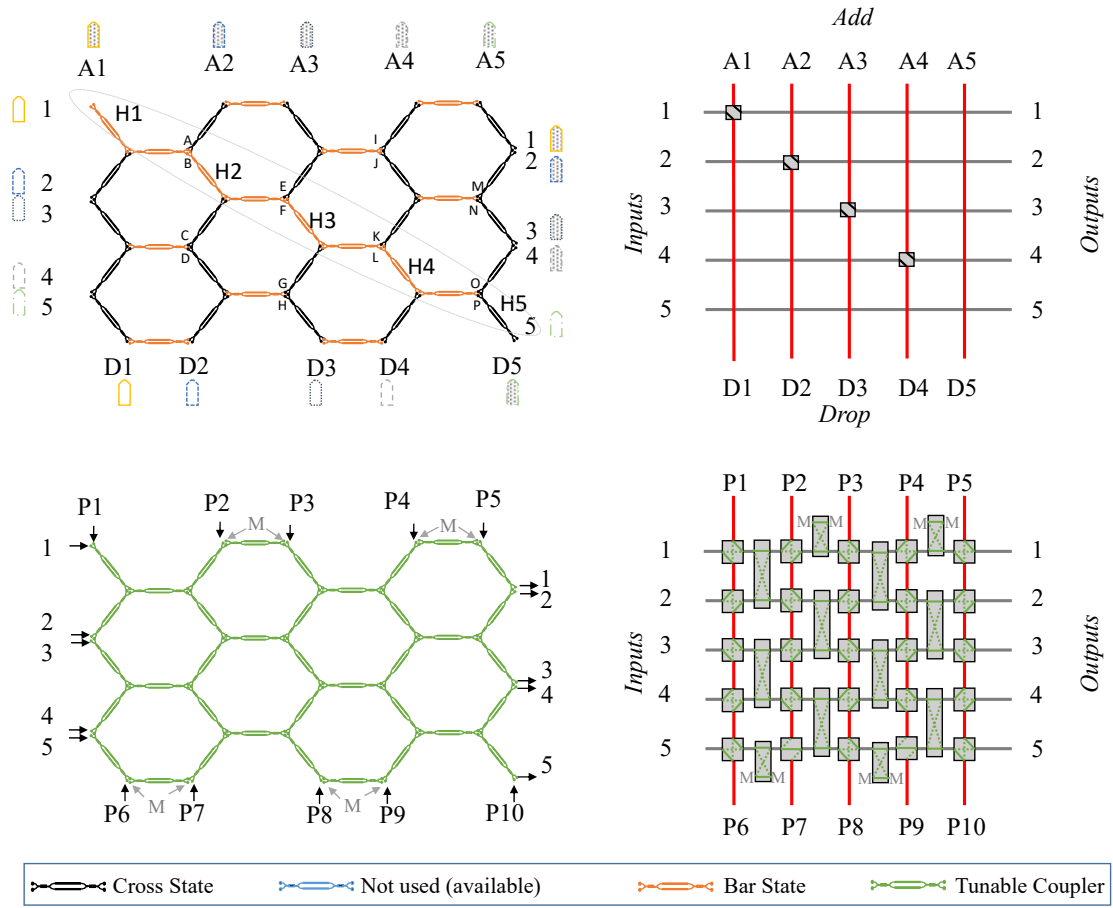

Figure 12. Waveguide mesh settings for channel management application: (up) Add-drop configurations for channels 1-4. Channel 5 bypasses the device. (Lower) fully reconfigurable channel management station that allows channel broadcasting, add/drops, channel combinations and demultiplexing. M: Signal monitoring points.

## 6. SUMMARY, CONCLUSIONS AND FUTURE DIRECTIONS

We have reviewed the recent advances reported in the field of integrated photonic waveguide meshes, both from the theoretical as well as from the experimental point of view showing how these devices can be programmed to implement both traditional signal processing structures, such as finite and infinite impulse response filters, delay lines, beamforming networks as well as more advanced linear matrix optics functionalities. It is in this latter topic, where the true potential of 2D integrated waveguide meshes for the implementation of complex configurations, supporting linear matrix transformations, needs still to be unleashed. The first results indicate that these structures can be able to support basically any *N*x*N* unitary transformation by emulation of previously reported multiport interferometers. Yet the 2D structure might also provide their own implementation of the above transformations. Future work should be directed towards this exciting area of research.

## REFERENCES

[1] Miller, D. A.B., "Self-configuring universal linear optical component, " Photonics Research, 1(1) 1–15 (2013).
[2] Miller, D. A. B., "Self-aligning universal beam coupler, " Optics Express, 21(15), 6360-6370 (2013).
[3] Graydon, O., "Birth of the programmable optical chip, " Nature Photonics, 10(1) 1 (2016).

[4] Zhuang, L., Roeloffzen, C. G. H., Hoekman, M., Boller K.-J. and Lowery, A. J., "Programmable photonic signal processor chip for radiofrequency applications, " Optica, 2 (10) 854-859, (2015).

[5] Pérez, D., Gasulla, I., Capmany J. and Soref, R. A., "Reconfigurable lattice mesh designs for programmable photonic processors, " Optics Express, 24(11), 12093-12106, (2016).

[6] Capmany, J., Gasulla, I. and Pérez, D., "Microwave photonics: The programmable processor," Nature Photonics, 10(1) 6-8 (2016).

[7] Pérez, D. et al., "Multipurpose silicon photonics signal processor core, " Nature Communications, 8, 636, (2017).

[8] Pérez, D. et al., "Silicon Photonics Rectangular Universal Interferometer, " Lasers and Photon. Rev, 11 1700219 (2017).

[9] Ribeiro, A., Ruocco, A., Vanacker L. and Bogaerts, W., "Demonstration of a 4 × 4-port universal linear circuit," Optica 3, 1348-1357 (2016)

[10] Anoni, A. et al., "Unscrambling light—automatically undoing strong mixing between modes," Light Science and Applications 6 e17110 (2017).

[11] Capmany, J., Gasulla, I. and Pérez, D.,"Towards Programmable Microwave Photonics Processors, " IEEE J. Lightwave Tech. in press (2018).

[12] Chen, L.-N., Hall, E., Theogarajan, L. and Bowers, J., "Photonic switching for data center applications," IEEE Photon. J. 3, 834–844 (2011).

[13] Miller, D.A.B., "Silicon photonics: Meshing optics with applications," Nature Photonics, 11(6) 403-404 (2017).

[14] Thomas-Peter, N. et al., "Integrated photonic sensing," New J. Phys. 13, 055024 (2011)

[15] Carolan, J. et al., "Universal linear optics," Science 349, 711 (2015).

[16] Miller, D.A.B., "Perfect optics with imperfect components," Optica 2, 747–750 (2015).

[17] Harris, N.C. et al., "Quantum transport simulations in a programmable nanophotonic processor, " Nature Photonics 11(6) 447-452 (2017).

[18] Metcalf, B. J. et al., "Multiphoton quantum interference in a multiport integrated photonic device, " Nat. Commun. 4, 1356 (2013).

[19] Peruzzo, A., Laing, A., Politi, A., Rudolph, T. and O'Brien, J. L., "Multimode quantum interference of photons in multiport integrated devices," Nat. Commun. 2, 224 (2011).

[20] Shen, Y. et al., "Deep learning with coherent nanophotonic circuits," Nature Photonics 11(6) 441-446 (2017).

[21] Pérez, D. Integrated Microwave Photonic Processors using Waveguide Mesh Cores, PhD Thesis, Universitat Politècnica de València, (2017).

[22] Reck, M., Zeilinger, A., Bernstein, H. J. and Bertani, P., "Experimental realization of any discrete unitary operator, " Phys. Rev. Lett. 73, 58–61 (1994).

[23] Clements, W. R. et al., "Optimal design for universal multiport interferometers," Optica 3, 1460-1465 (2016).

[24] Micó, G. et al., , "C-band linear propagation properties for a 300 nm film height Silicon Nitride photonics platform, " European Conference on integrated optics (ECIO) 2017, Eindhoven, the Netherlands., (2017).

[25] Madsen, C. K. and Zhao, J. H., Optical filter design and analysis: A signal processing approach, John Wiley & Sons, Inc., (1999).

[26] Celo, D. et al., "32x32 silicon photonic switch," in OptoElectronics and communications conference (OECC) & IEEE Photonics in Switching (PS), 2016.

[27] Sheng, Z. et al., "A Compact and low-loss MMI coupler fabricated with CMOS Technology," IEEE Photonics Journal, vol. 4, nº6, pp. 22272-2277, 2012.

[28] Cong, G. W. et al., "Demonstration of a 3-dB directional coupler with enhanced robustness to gap variations for silicon wire waveguide." Optics express, vol. 22, nº 2, pp. 2051-2059, 2014.

[29] Harris, N. C. et al., "Efficient, compact and low loss thermos-optic phase shifter in silicon," Optics Express, vol. 22, nª9, pp. 10487-10493, 2014.

**ACKNOWLEDGMENTS**

This research was supported by the ERC Advanced Grant 741415 UMWP-CHIP, the COST Action CA16220 EUWMP, the Spanish MINECO Projects TEC2014-60378-C2-1-R and the Generalitat Valenciana PROMETEO Project 2017/103. Ivana Gasulla acknowledges the Spanish MINECO Ramon y Cajal program RYC-2014-16247.